\documentclass[12pt]{article}
\usepackage{times}
\newenvironment{sciabstract}{%
\begin{quote} \bf}
{\end{quote}}
\usepackage{scicite}

\usepackage[margin=1in]{geometry}
\usepackage{multicol,caption}
\usepackage[usenames,dvipsnames]{color}
\usepackage{amssymb}
\usepackage{amsmath,amssymb,amsfonts}
\usepackage{graphicx}
\usepackage{graphics} 
\usepackage{cancel}
\usepackage{amsmath}
\usepackage{epstopdf}
\usepackage{fixltx2e}
\usepackage{url}
\usepackage{bm}
\usepackage{subfigure}
\usepackage{dcolumn}
\usepackage[hidelinks]{hyperref}
\usepackage{setspace}
\usepackage{wrapfig}

\newcounter{lastnote}
\newenvironment{scilastnote}{%
\setcounter{lastnote}{\value{enumiv}}%
\addtocounter{lastnote}{+1}%
\begin{list}%
{\arabic{lastnote}.}
{\setlength{\leftmargin}{.28in}}
{\setlength{\labelsep}{.5em}}}
{\end{list}}

\graphicspath{{./imgs/}{./}}

\title{Probing the Bose-Glass--Superfluid Transition using Quantum Quenches of Disorder}

\author{C. Meldgin,$^{1}$ U. Ray,$^{1}$ P. Russ,$^{1}$ D. Ceperley,$^{1}$ and B. DeMarco$^{1\ast}$\\
\\
\normalsize{$^{1}$University of Illinois at Urbana-Champaign,}\\
\normalsize{1110 W Green St, Urbana, IL 61801, USA}\\
\normalsize{$^\ast$To whom correspondence should be addressed; E-mail:  bdemarco@illinois.edu.}
}

\date{\today}                                           

\begin{document}

\baselineskip24pt
\maketitle

\begin{sciabstract}
We probe the transition between superfluid and Bose glass phases using quantum quenches of disorder in an ultracold atomic lattice gas that realizes the disordered Bose-Hubbard model.  Measurements of excitations generated by the quench exhibit threshold behavior in the disorder strength indicative of a phase transition.  Ab-initio quantum Monte Carlo simulations confirm that the appearance of excitations coincides with the equilibrium superfluid--Bose-glass phase boundary at different lattice potential depths.  By varying the quench time, we demonstrate the disappearance of an adiabatic timescale compared with microscopic parameters in the BG regime.
\end{sciabstract}

The dynamics resulting from tuning, or quenching, quantum matter across a phase transition provide fundamental insights into the nature of many-particle systems \cite{Polkovnikov}.  One approach to understanding this problem is the Kibble-Zurek (KZ) scenario \cite{Kibble,Zurek,DziarmagaReview}, which links the dynamical generation of excitations as equilibrium is disrupted during a quench to the critical exponents of the equilibrium phase transition.  Despite the prevalence of disorder in quantum matter, little is known about how disorder influences non-equilibrium dynamics and affects the KZ paradigm in closed quantum systems \cite{gogolin}.  Understanding dynamic and non-equilibrium properties of disordered quantum materials is of paramount importance to applications such as quantum annealing of disordered spin systems to benchmark adiabatic quantum computing \cite{Santoro29032002}.  In classical systems, disorder can have a profound impact and lead to the formation of glasses, which display quench phenomena such as aging, experimentally inaccessible equilibration times, and relaxation dynamics that are unrelated to equilibrium states \cite{young}.  Whether or not analogous behavior occur in strongly interacting disordered quantum systems is an open question.

Here, we use a quantum quench of disorder in an ultracold lattice gas to probe the superfluid--Bose-glass (SF--BG) quantum phase transition (Fig. 1).  We show that the appearance of excitations generated by the quench is determined by the ground-state phase diagram via comparisons to quantum Monte Carlo (QMC) simulations.  By varying the quench time, we demonstrate that this sensitivity to the many-particle ground state occurs despite the disappearance of an adiabatic timescale (compared with microscopic parameters) in the BG regime.  To make these measurements, we create an atomic realization of the three-dimensional disordered Bose-Hubbard model (DBHM) using ultracold $^{87}$Rb atoms trapped in a disordered optical lattice \cite{White}.  The DBHM is a paradigm for strongly correlated and disordered bosonic systems, such as $^4$He in disordered substrates like aerogels, disordered Josephson-junction arrays, and long-wavelength properties of superconducting electron pairs \cite{Fisher}.  In the DBHM, a strongly interacting SF undergoes a quantum phase transition into a BG when subjected to disorder.  The BG phase exhibits the peculiar property of lacking long-range order while possessing infinite superfluid susceptibility \cite{Fisher}, and it is therefore viewed as a gapless insulator with finite compressibility that arises from the presence of quasi-condensates, or SF puddles, embedded in an insulating background. Disordered ultracold atom gases have been used to indirectly measure the SF--BG transition via transport and coherence measurements in 1D \cite{PhysRevLett.107.145306} and 3D \cite{Pasienski} disordered lattices and in 1D quasi-periodic lattices \cite{Inguscio,Inguscio2}.

\begin{figure*}[h!]
\centering
\includegraphics[width=0.7\textwidth]{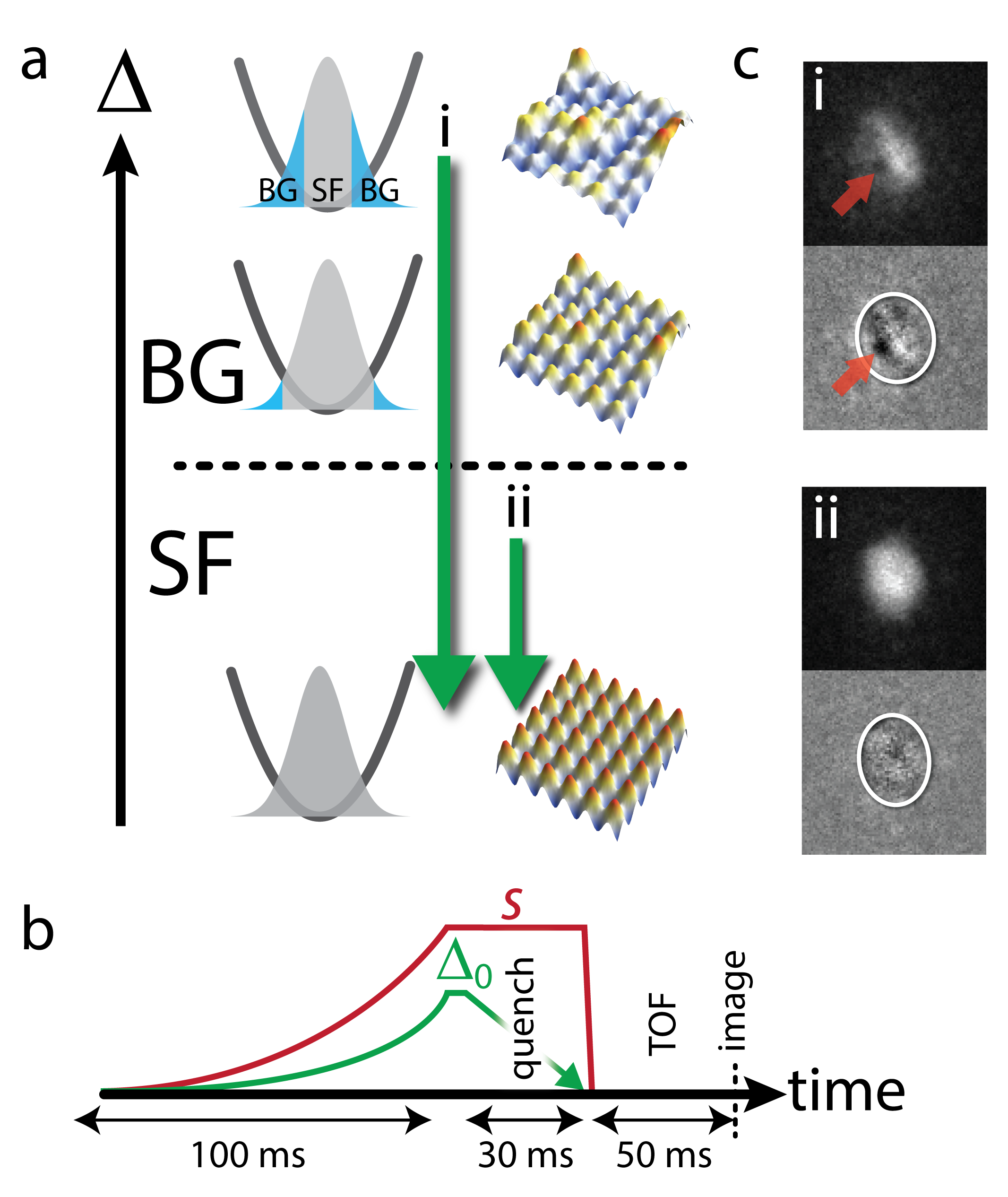}
\label{fig1}
\caption{(a) Schematic phase diagram of the DBHM and quench.  The gas is quenched from the BG to the SF regime by rapidly reducing (green arrow) the disorder strength $\Delta$ from $\Delta_0$ to zero at fixed $U/t$, which is determined by the lattice potential depth $s$. Equilibrium configurations and the disordered lattice potential (false color) are shown at three values of $\Delta$.  For sufficiently high $\Delta$, BG (blue) and SF (light gray) phases coexist in the trap. (b) Time sequence for the measurement.  The lattice potential depth and disorder strength are shown using red and green lines.  (c) Equilibration is disrupted during the quench and excitations are produced, which are measured in TOF images (grayscale).  Images are shown for $\Delta_0\approx0.5$~$E_R$ (i) and $\Delta_0=0$ (ii) at $s=12$~$E_R$. The white ellipse marked on the residual from a fit to the image marks the fitted TF radius.  For sufficiently high disorder, excitations such as vortices are apparent (red arrow) after the quench, while smooth profiles are obtained at low $\Delta_0$.}
\end{figure*}

In our experiment, we prepare a gas consisting of $\left(27\pm2\right)\times 10^3$ atoms cooled to a few nanoKelvin in a harmonic optical dipole trap.  A disordered cubic optical lattice formed from pairs of counter-propagating $\lambda=812$~nm laser beams and a 532~nm optical speckle field is superimposed on the gas \cite{SM}.  The atoms experience a potential energy shift proportional to the speckle intensity, which varies randomly in space, leading to disorder in the Hubbard tunneling $t$, interaction $U$, and site occupation $\epsilon$ energies.  The DBHM we realize is characterized by the Hamiltonian
\begin{equation}
H = -\sum_{<ij>}t_{ij} \hat{b}^\dagger_i \hat{b}_j  + \sum_i (\epsilon_i - \mu) \hat{n}_i
+ \frac{1}{2} \sum_i U_i \hat{n}_i (\hat{n}_i-1)  +\frac{1}{2}\sum_i m \omega^2 r_i^2 \hat{n}_i,
\end{equation}
where $i$ and $j$ index the lattice sites and $\langle\rangle$ indicates that tunneling occurs only between adjacent sites.  In Eq. 1, $\hat{n}_i$ is the number of particles on site $i$, $\hat{b}_i$ ($\hat{b}_i^\dagger$) removes (adds) a particle from site $i$, $m$ is the atomic mass, $\omega$ is the geometric mean of the trap frequencies, $r_i$ is the distance to the center of the trap, and $\mu$ is the chemical potential. We measure all energies in terms of the recoil energy $E_R=h^2/2m\lambda^2\approx 170$nK~$k_B$. The distribution of the Hubbard parameters, which are broadened around the values for the uniform system, are precisely known \cite{White,Zhou}.  The strength of the disorder is characterized by the average potential energy $\Delta$ associated with the speckle, which is approximately equal to the standard deviation of the distribution of site occupation energies.  The lattice potential depth $s$ (which controls $U/t$) and $\Delta$ are independently adjusted by tuning the power of the lattice laser and 532~nm light.  The range of $s$ we sample in this work corresponds to a strongly correlated, quantum depleted SF for $\Delta=0$.

We probe the BG--SF  transition by measuring the amount of excitation produced by quenching $\Delta$ at fixed $s$.  The disorder strength is linearly ramped from an initial value of $\Delta_0$ to zero in 30 ms (Fig 1b), which is slow enough to avoid creating excitations solely via the time variation of the spatially inhomogeneous disorder potential \cite{SM}.  Based on general arguments regarding the phase diagram in untrapped systems, the BH phase will appear in the low-density edge of the gas for sufficiently high $\Delta_0$ \cite{Fisher}.  For stronger disorder, the BG--SF boundary moves inward, encompassing more of the atoms.  Excitations produced by the quench are measured using time-of-flight (TOF) imaging.  By imaging after a long (50~ms) period of free expansion, vortices and other excitations are transformed into modulations of the density profile and the measured optical depth ($OD$).  These excitations are visible in the characteristic images shown in Figs. 1c and 2b.  For low $\Delta$, the density profile after the quench and TOF is smooth, while for high $\Delta$, features consistent with vortices are present.

To quantitatively characterize the amount of excitation present after the quench, we measure
\begin{equation}
\tilde{\chi}^2 =\sum_{ij}\frac{(\mathcal{O}_{ij}-f_{ij})^2}{f_{ij}}/\sum_{ij}\mathcal{O}_{ij},
\end{equation}
where $\mathcal{O}_{ij}$ is the measured $OD$ at the pixel indexed by $i$ and $j$ within a mask set by a smooth fitting function $f$ that is the combination of a Thomas-Fermi (TF) profile and a gaussian (which is approximately the equilibrium SF distribution).  This method was previously used to observe the quantum KZ effect by measuring excitations generated via a quench between MI and SF states in a ``clean" lattice \cite{Chenzo}.  Data for $s=11$~$E_R$ and $\Delta\approx0$--1~$E_R$ are shown in Fig. 2a.  It is apparent that excitations are not generated by the quench until a threshold disorder strength is crossed, above which $\tilde{\chi}^2$ increases approximately linearly with $\Delta_0$.  Similar threshold behavior is observed for all $s$ we sample in this work.

\begin{figure*}[h!]
\centering
\begin{tabular}{lr}
\includegraphics[width=.85 \textwidth]{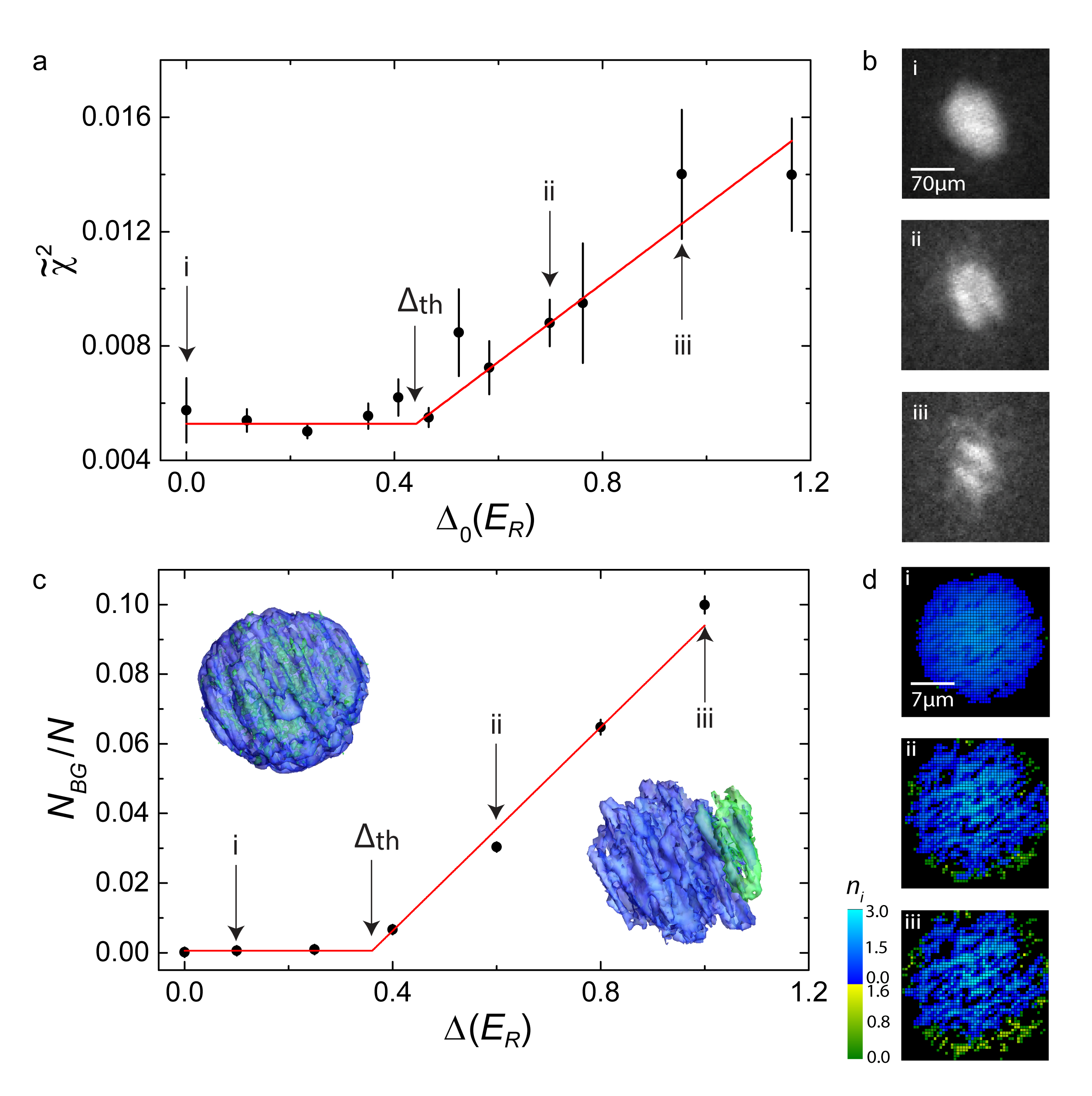}
\end{tabular}
\caption{(a)-(b) Results of quench measurements. (a) The observed $\tilde{\chi}^2$ as $\Delta_0$ is varied at $s=11$~$E_R$ and the piecewise linear fit (red line) used to extract the threshold disorder $\Delta_{th}$ are shown. The error bars show the standard error in the mean for the 6--12 measurements averaged at each $\Delta_0$.  TOF images obtained after the quench are shown for $\Delta_0=0$ (i), $\Delta_0=0.7$ (ii), and $\Delta_0=0.95$~$E_R$ (iii). (c)-(d) Results from QMC simulations.  (c) The upper bound $N_{BG}/N$ on the BG fraction is shown as a function of $\Delta$ for $s=11$~$E_R$.  The error bars show the standard error in the mean for the QMC statistical noise. The insets are three-dimensional contour plots of the highest (blue) and second highest (green) occupation eigenfunction of the single-particle density matrix for $\Delta=0.05$ (upper left) and $\Delta=1$~$E_R$ (lower right). (d) Density slices through the trap center are shown for $\Delta=0.1$ (i), $\Delta=0.6$ (ii), and $\Delta=1$~$E_R$ (iii).  The blue (green) regions are the SF (BG) domains, and the color bar shows the average number of particles on each site.}
\end{figure*}

This threshold behavior is generally associated with a phase transition and the quantum KZ effect \cite{Kibble,Zurek,DziarmagaReview}. In the KZ scenario, an adiabatic transition from a phase disordered (e.g., BG) to an ordered (e.g., SF) state is impossible because of diverging characteristic length and time scales.  Dynamically traversing a quantum phase transition by tuning (or quenching) a Hamiltonian parameter such as $\Delta$ necessarily leads to the formation of domains and excitations such as vortices that persist even after the transition is crossed \cite{Polkovnikov}.  In this case, the excitations occur when the BG state is present in the gas and the SF--BG transition is crossed as the disorder strength is reduced during the quench.

To connect the observed threshold disorder with the SF--BG transition, we carry out exact QMC simulations of the equilibrium system using the same trap and lattice parameters, atom number, and speckle disorder as in the experiment \cite{SM}.  For trap-free geometries in the thermodynamic limit, the BG is characterized by a vanishing superfluid order parameter and non-zero compressibility \cite{Pollet2}. In contrast, the trapped system we consider exhibits domains corresponding to SF and BG phases that we distinguish using the spatial extent of the condensate. The condensate is identified as the macroscopic occupation of a single-particle eigenstate that we can obtain from the single particle density matrix $\rho_1 \equiv \sum_{ij} \langle \hat{b}^\dagger_i \hat{b}_j \rangle$ \cite{Onsager,Leggett,Ceperley,Ray}.

For clean systems ($\Delta = 0$) and $U/t < 29.34\pm0.02$ at ultra low temperatures ($k_BT/12t \ll 1$), a single condensate extends throughout the system that coincides with local superfluid density order parameter \cite{SM}. As $\Delta$ is increased, this behavior changes and the extent of the macroscopic condensate shrinks, leaving behind regions devoid of coherence with it. Since the SF--BG transition is of the continuous type, phase coexistence is forbidden \cite{Vojta}, and we identify these regions as BG.  To illustrate this behavior, we show the two highest occupation eigenfunctions of $\rho_1$ for $s=11$~$E_R$ and $\Delta=0.05$ and 1~$E_R$  in Fig. 2c. At low $\Delta$, all single-particle states are spatially overlapped with the SF domain, and the second highest occupied state results from interaction-induced quantum depletion. For sufficiently high $\Delta$, however, this extended state is replaced by a spatially localized mode that corresponds to a non-macroscopic and locally coherent superfluid puddle characteristic of the BG phase.

To compare with the measurements, we compute the BG fraction $N_{bg}/N$ as the fraction of atoms in regions without a macroscopic condensate present.  This estimate is an upper bound at non-zero temperature because of thermal excitation.  As shown in Fig. 2d, the BG as defined by this criterion emerges at the edge of the gas and grows in extent and number as $\Delta$ is increased.  Typical behavior for $N_{bg}/N$ at $s=11$~$E_R$ as $\Delta$ is varied is shown in Fig. 2c.  Similarly to the amount of excitation created by the quench in the experiment, $N/N_{bg}$ is only non-zero above a threshold disorder, above which it increases approximately linearly with $\Delta$.

We construct the SF--BG phase diagram shown in Fig. 3 by estimating the threshold disorder $\Delta_{th}$ for generating excitations in the experiment and for BG to appear in QMC simulations using a piecewise-linear fit to data such as those shown in Fig 2. The fitting function assumes constant behavior for disorder strengths less than $\Delta_{th}$ and linearly increasing behavior characterized by the free parameters $\Delta_{th}$ and a slope for disorder strengths greater than $\Delta_{th}$.  Several important features of the phase diagram are evident.  The threshold disorder $\Delta_{th}$ is weakly dependent on $s$, and the QMC and experimental results agree within the 40\% systematic uncertainty in $\Delta_0$; there are additional systematic and statistical uncertainties arising from finite temperature and disorder averaging \cite{SM}.  This agreement---which demonstrates that the quench dynamics and production of excitations in this strongly disordered system are sensitive to the ground-state, equilibrium phases---supports the quantum KZ scenario. Furthermore, the observed threshold behavior cannot be explained by mean field theory, which predicts that a BG appears for infinitesimal disorder \cite{Hofstetter}.  Finally, the decrease in $\Delta_{th}$ at higher $s$ (i.e., larger $U/t$), which cannot be accounted for by general classical percolation mechanisms, implies that interactions facilitate the transition from SF to BG.

\begin{figure*}[h!]
\centering
\includegraphics[width=.7 \textwidth]{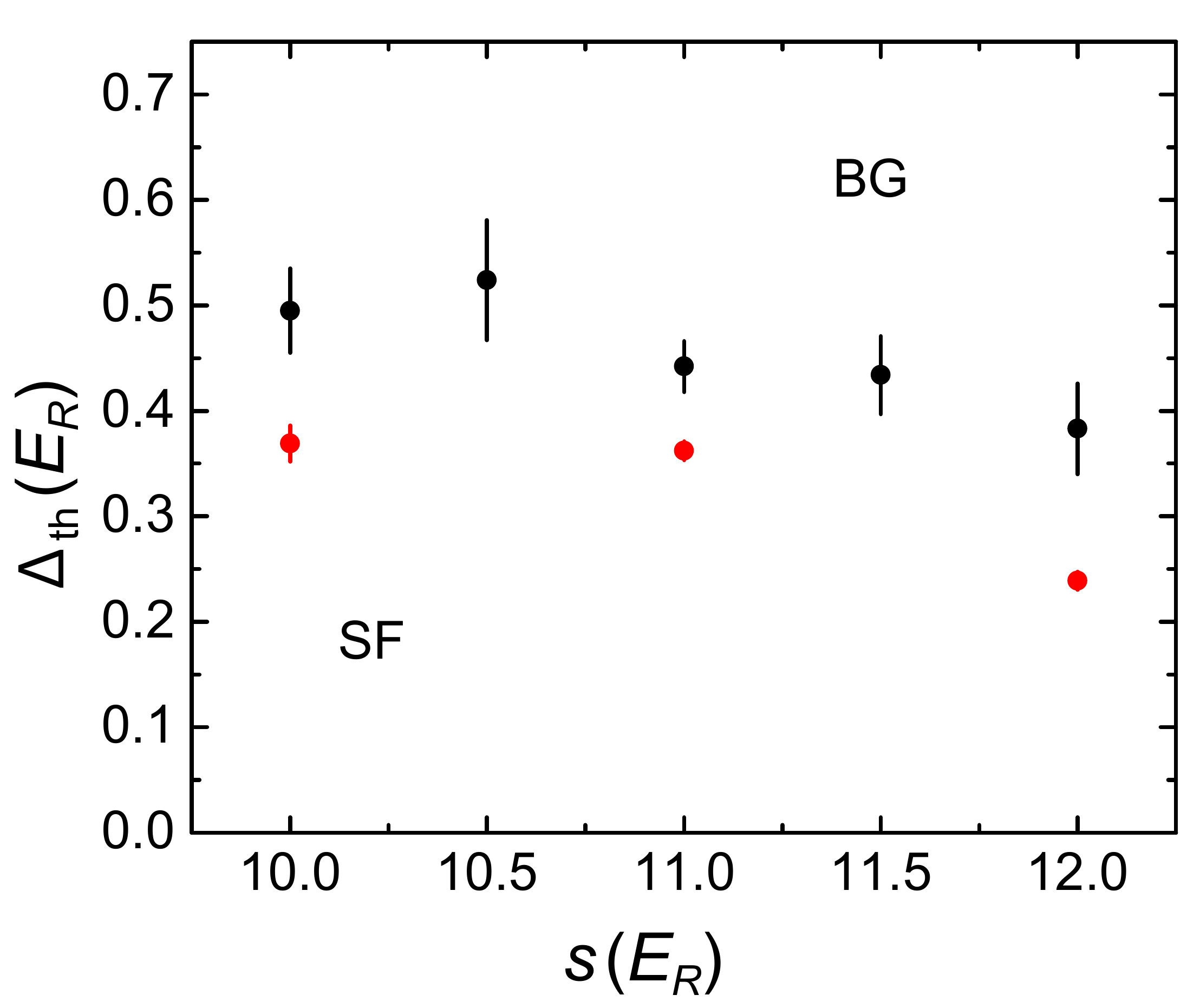}
\caption{Phase boundary between SF and BG regimes.  Every point is the result from a piece-wise linear fit to data at fixed $s$ (such as those shown in Figs. 2a and c), and the error bars show the fit uncertainty. The black squares are the experimentally determined values of $\Delta_{th}$ from quench measurements, and the red circles are the QMC simulation results.}
\end{figure*}

We explore the dynamical timescale of the SF--BG transition by varying the quench time.  In the KZ scenario for clean systems, the amount of excitation and heat produced during a quantum quench typically display power-law dependence on the quench time \cite{DziarmagaReview}. The knowledge of how this changes in disordered systems is limited to one-dimensional spin chains, which show logarithmic behavior \cite{PhysRevB.74.064416,PhysRevB.76.144427}.  In our experiment, the dependence of $\tilde{\chi}^2$ on the quench time $\tau_q$ when the SF--BG boundary is crossed is too weak to detect.  Typical data are shown in Fig. 4 at $s=10$~$E_R$ for the SF domain at a disorder strength just below $\Delta_{th}$ ($\Delta_0=0.35$~$E_R$) and for the BG regime ($\Delta_0=1$~$E_R$).  To avoid complications from decay of excitations during the quench, we determine the amount of excitation by measuring the temperature $T$ of the gas after allowing rethermalization in the trap for 150 ms \cite{SM}.  We show the fractional deviation in the temperature $\left(T-T_0\right)/T_0$ in Fig. 4, where $T_0$ is the temperature of the gas without disorder applied, in order to normalize heating from the lattice laser light.  For the SF domain, a characteristic timescale for excitations to occur is evident.  The data fit well to a decaying exponential function with a time constant $4.4\pm1.6$~ms.  Quenches with $\tau_q$ much longer than this time do not produce excitation.  In contrast, in the BG regime, an adiabatic timescale is absent for $\tau_q$ up to 140~ms, which is approximately an order of magnitude longer than the tunneling time $h/t\approx 15$~ms: the slowest microscopic timescale present in the DBHM Hamiltonian.  We cannot explore larger $\tau_q$ because heating from the lattice light results in a loss of the signal-to-noise ratio.

\begin{figure*}[h!]
\centering
\includegraphics[width=.7 \textwidth]{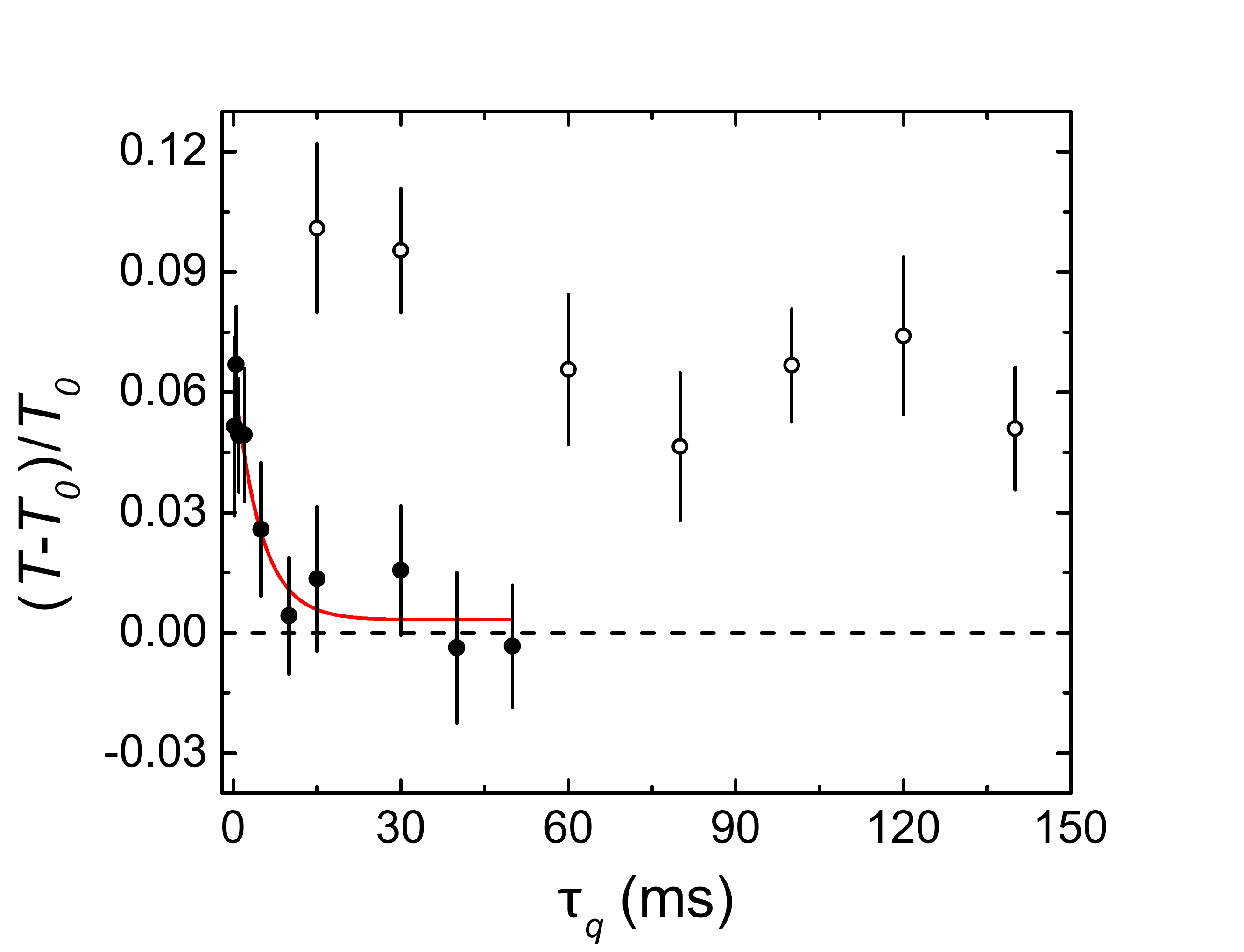}
\caption{Fractional change in temperature of the gas after quenches with different quench times $\tau_q$.  The filled circles are for the SF phase with $\Delta_0=0.35$~$E_R$, which is smaller than the threshold disorder $\Delta_{th}$. The open circles are for $\Delta_0=1$~$E_R$, which is greater than $\Delta_{th}$ and corresponds to the BG regime.  The error bars shown the standard error in the mean for the 8--15 measurements averaged for each point.  The solid line is a fit to an exponential decay.}
\end{figure*}

The emergence of an equilibration time much longer than microscopic timescales is reminiscent of glassy behavior in disordered classical systems \cite{young}.  Our understanding of dynamics in the DBHM for two and three dimensions is limited \cite{PhysRevB.86.214207}, since direct simulation is intractable for experimentally relevant numbers of particles.  Whether the long timescale we observe in the BG phase is associated with critical phenomena or if is is connecting solely with disorder-induced glassiness is thus an open question.  More work, such as measurements of how correlations change during the quench \cite{Navon09012015}, is needed to clarify how the KZ scenario is altered by disorder.

\bibliography{DisorderQuenchPaper_bd}
\bibliographystyle{Science}

\begin{scilastnote}
\item The authors acknowledge funding from the National Science Foundation and the Army Research Office. Computation time was provided by XSEDE resources at TACC (Texas) and INCITE resources at Oak Ridge National Laboratory.  C. Meldgin and U. Ray contributed equally to this work.
\end{scilastnote}

\end{document}